\documentclass[twocolumn, prb,preprintnumbers,amsmath,amssymb,floatfix,superscriptaddress]{revtex4-2}
\usepackage[linktocpage,bookmarksopen,bookmarksnumbered]{hyperref}

\usepackage{dcolumn}

\usepackage{amsmath,graphics,epsfig,color,verbatim,ulem}
\usepackage{braket}

\usepackage{color}
\usepackage{soul}

 \newcommand{\vk}{{\mathbf{k}}}

\usepackage{textcomp}

\usepackage{tabularx,ragged2e,booktabs,caption}
\newcolumntype{C}[1]{>{\Centering}m{#1}}

\usepackage{multibib}

\begin{document}

\setlength{\pdfpageheight}{\paperheight}
\setlength{\pdfpagewidth}{\paperwidth}

\title{Role of orbital selectivity on crystal structures and electronic states in BiMnO$_3$ and LaMnO$_3$ perovskites}

\author{Gheorghe Lucian Pascut}
\affiliation{MANSiD Research Center and Faculty of Forestry, Stefan Cel Mare University (USV), Suceava 720229, Romania}
\affiliation{Center for Materials Theory, Department of Physics $\&$ Astronomy, Rutgers University, Piscataway, NJ O8854, USA}

\author{Kristjan Haule}
\affiliation{Center for Materials Theory, Department of Physics $\&$ Astronomy, Rutgers University, Piscataway, NJ O8854, USA}
%




\date{\today}
\begin{abstract}
Correlated oxides, such as BiMnO$_3$ and LaMnO$_3$, show complex interplay of electronic correlations and crystal structure exhibiting multiple first order phase transitions, some without a clear order parameter. The quantitative theoretical description of this temperature dependent electronic-structural interplay in the vicinity of a Mott transition is still a challenge. Here we address this issue by simultaneously considering both structural and electronic degrees of freedom, within a self-consistent density functional theory with embedded dynamical mean field theory. Our results show the existence of novel electronic states characterized by coexistence of insulating, semi-metallic and metallic orbitals. This state is in agreement with resonant X-ray scattering. We also show that electronic entropy plays a decisive role in both electronic and structural phase transitions. By self-consistent determination of both, the electronic state and the corresponding crystal structure, we show that the temperature evolution of these phases can be quantitatively explained from first principles, thus demonstrating the predictive power of the theoretical method for both the structural and the electronic properties.
\end{abstract}
\maketitle

Manganese oxides with perovskite-type structure host a variety of interesting phenomena, among others the
colossal magnetoresistance~\cite{colossal_magnetoresistance_glp_DAGOTTO20011, colossal_magnetoresistance_glp1_0953-8984-9-39-005}, insulator to metal transition~\cite{Metal_insulator_transitions_glp_RevModPhys.70.1039}, ferroelectricity coexisting with long range magnetic order~\cite{Multiferroics_glp_KHOMSKII20061}, and orbital and charge ordering~\cite{Orbital_Physics_glp_Tokura462}. The theoretical description of these functionalities is challenging, as the electronic correlations play an essential role, and the interplay  between the electronic, orbital, and magnetic degrees of freedom with the crystal structure makes these materials extremely sensitive to small changes in external parameters, such as electric voltage, magnetic field, and temperature. However, this high sensitivity is a useful property, which is explored in development of new electronic devices.

The theoretical methods for ab-initio description of strong correlations in solids have been recently developed, and among them the Density Functional Theory with Embedded Dynamical Mean Field Theory (DFT+eDMFT)~\cite{webpage,Haule_prb10,JPSJ_Haule} has demonstrated predictive power in many classes of correlated materials. For example, it was predicted that the magnetic moment in plutonium metal is screened at $800~$K with valence fluctuations~\cite{Shim_nat07}, which was subsequently verified by neutron scattering experiment~\cite{Janoscheke1500188}. 
An order of
magnitude enhanced electron-phonon coupling in FeSe superconductor as
compared to the DFT result was predicted
in Ref.~\onlinecite{PhysRevB.89.220502}, and subsequently verified by
pioneering experiment using combination of time-resolved photoemission
spectroscopy coupled with x-ray
diffraction~\cite{Gerber71}. 
Similarly, prediction of large fluctuating moments in iron superconductors and their energy distribution~\cite{Yin-nm11} have been later verified by neutron scattering~\cite{Pengcheng}. The prediction of isostructural metal-insulator transition under pressure in FeO has been confirmed in Ref.~\onlinecite{PhysRevLett.108.026403}. Novel type of long ranged order with very high multipole-ordering was proposed to explain the hidden order in URu$_2$Si$_2$~\cite{Haule_np09} which was confirmed by Raman scattering experiment~\cite{Kung1339}. Finally, structural relaxations by eDMFT method have been recently developed~\cite{Forces_DMFT_glp_PhysRevB.94.195146} and it was demonstrated on the example of NdNiO$_3$ ~\cite{RNiO3_glp_Scientific_Reports} and Fe$_2$Mo$_3$O$_8$~\cite{FeMo} that they can predict small changes of crystal structures near the Mott metal-insulators transition, the region which is challenging for other ab-initio methods.

\begin{figure*}[bht]
\includegraphics[width=0.96\linewidth]{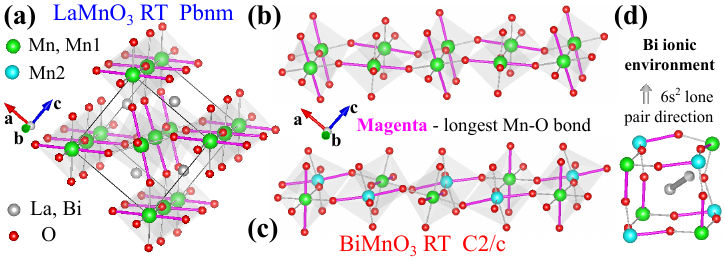}
\caption{(Color online)  \textbf{Room temperature (RT) structural properties for LaMnO3 and BiMnO3.} (a) schematic view of the crystal structure for LaMnO$_3$. The longest Mn-O bond lengths are colored by magenta;  (b) displays the octahedra distortion pattern in the (101) plane; (c) distortions in BiMnO$_3$ in which Mn$_1$ ions show similar pattern as in LaMnO$_3$, while in Mn$_2$ ions the long-bond distortions are confined to (101) plane; (d) schematic view of Bi environment within a perovskite like crystal structure; the gray arrow, pointing towards the face of the  Mn$_2$O$_6$ octahedra, shows the direction of the 6s$^2$ lone-pair electronic cloud within the Bi ionic environment~\cite{LP_JT_2_glp_theory_10.1021_cm010090m}. The presence of the 6s$^2$ lone-pair is responsible for the appearance of two inequivalent Mn sites. Drawings were produced by Vesta~\cite{Vesta_Momma_db5098}.} 
\label{fig2}
\end{figure*}
 In this paper we study AMnO$_3$ class of materials with perovskite structure-type to understand the role that correlations play in generating their rich phase diagrams~\cite{PD_1_glp_PhysRevLett.106.066402, OO_2017_glp_PhysRevB.95.174107,PD_2_glp_PhysRevB.80.184426,
PD_3_glp_Alexe_Belik2009}. 
We compare and contrast LaMnO$_3$ with BiMnO$_3$, which have attracted considerable interest due to their giant magnetoresistance phenomena in doped LaMnO$_3$~\cite{colossal_magnetoresistance_glp_DAGOTTO20011}, and potential multiferroic property in BiMnO$_3$ thin films ~\cite{Multiferroics_BiMnO3_glp_BELIK201232,gapTHeory4_BiMnO3_PhysRevB.91.184113}, respectively.

The high temperature (HT) phase ($T>800\,$K) of both LaMnO$_3$ and BiMnO$_3$ is perovskite-type structure with orthorhombic symmetry~\cite{BiMnO3_HT_LP1_glp_PhysRevB.67.180401, BiMnO3_HT_LP2_glp_B818645F, LaMnO3_CS_glp_PhysRevB.57.R3189} (see Fig.~\ref{fig2}a) with similar bad-metallic conductivity~\cite{BiMnO3_HT_LP1_glp_PhysRevB.67.180401,LaMnO3_HT_P1_rezis_glp,LaMnO3_HT_P2_r2_glp, LaMnO3_HT_P1_rezis3_glp}.
At room temperature (RT) both compounds are insulating~\cite{BiMnO3_HT_LP1_glp_PhysRevB.67.180401, LaMnO3_INS_RT_glp_PhysRevB.64.180405} and in LaMnO$_3$ the orthorhombic distortions increases with decreasing temperature but the symmetry remains orthorhombic ~\cite{LaMnO3_CS_glp_PhysRevB.57.R3189},
while in BiMnO$_3$ the structure distorts to
monoclinic symmetry~\cite{MM_CS_BiMnO3_glp_JACS,RT_CS_BiMnO3_glp_PhysRevB.77.024111,RT_CS_BiMnO3_glp_RAMAN_PhysRevB.89.224107}.
An abrupt change of resistivity is observed in both compounds around the transition, with the high-temperature phase exhibiting nearly temperature independent resistivity, characteristic of a very bad metal.
The pattern of structural distorsions in both compounds is compatible with the long-range orbital ordering (OO)~\cite{LaMnO3_CS_glp_PhysRevB.57.R3189,MM_CS_BiMnO3_glp_JACS,OO_LaMnO3_Raman_PhysRevLett.92.097203,OO_proposals_IOP_glp_1367-2630-10-7-073021}, which has been validated by resonant X-ray scattering (RXS) experiments~\cite{LaMnO3_RXS_Colella_dm0001,LaMnO3_RXS_good_PhysRevLett.81.582, LaMnO3_RXS_PhysRevB.73.224112} (see Fig.~\ref{fig2}b and c). 
The theoretical description of this Mott-type transition is most often described within the phenomenological models of cooperative Jahn-Teller (JT) effect due to Mn $d^4$ configuration in octahedra, leaving a single electron in the $e_g$ shell~\cite{JT_1_glp_theory_PhysRevB.56.12154,JT_2_glp_theory_PhysRevB.82.045124,LP_JT_1_glp_theory_PhysRevB.82.245101,LP_JT_2_glp_theory_10.1021_cm010090m,LaMnO3_HT_P1_rezis_glp, OO_2017_glp_PhysRevB.95.174107,OO_potts_PhysRevB.79.174106}.  However, the two $e_g$ orbitals are not degenerate in the high-temperature metallic phase, therefore the conventional JT mechanism does not apply, and better explanation of the transition is called for.
Moreover, experimentally it was found that BiMnO$_3$ undergoes two phase transitions, a structural phase transition from orthorhombic to monoclinic symmetry takes place at $T=768\,$K~\cite{BiMnO3_HT_LP1_glp_PhysRevB.67.180401}, followed by another isostructural phase transition at $T=474\,$K~\cite{MM_CS_BiMnO3_glp_JACS,BiMnO3_HT_LP1_glp_PhysRevB.67.180401,BiMnO3_Mossbauer_Glazkova2016}. 
While orbital order is symmetry allowed at the first transition, RXS experiments showed that the situation is more complicated.~\cite{LaMnO3_RXS_PhysRevB.73.224112}  
There is a sharp mean-field type upturn 
of the forbidden Bragg peak intensity 
at $T=768\,$K followed by the saturation regime. Below the second transition ($T=474\,$K), there is a second sharp upturn of the intensity, signaling that the long range orbital order is well established only below the second phase transition ($T<474$). 
The forbidden Bragg peak intensity in the intermediate phase ($474\,$K$<T<768\,$K) saturates to $\approx 0.6$ of the room temperature intensity, which suggests that the intermediate phase has some type of partial order, but the nature of this partial order remains unknown. Both transitions are of first order, but the second isostructural transition does not have a clear order parameter.

Mott transition and the Jahn-Teller distortion in LaMnO$_3$ has been previously studied by Dynamical Mean Field Method using the minimal Hubbard model for the $e_g$ electron, while the $t_{2g}$ electrons were treated semi-classically, as a classical spin $S$ with random orientation.~\cite{Pavarini,Leonov} In this approach the effect of covalency is ignored, which was shown by X-ray absorption (XAS) to be important for interpretation of the XAS data~\cite{Xray2m}. Nevertheless, the metal-insulator transition was well captured by this minimal model approach~\cite{Pavarini,Leonov}. In Ref.~\onlinecite{Pavarini} the transition temperature to disordered state was somewhat underestimated, because the feedback effect of electronic correlations on crystal structure was ignored. This was partially corrected in Ref.~\onlinecite{Leonov} by allowing JT distortion, although the full structural relaxation was not attempted. Nevertheless much better estimate for the transition temperature was reported, despited the fact that the electronic entropy was ignored in the calculation of energy. On the other hand,  the complex phase diagram of BiMnO$_3$ has not been theoretically addressed before.

To understand the unusal Mott transition coupled with the partial orbital order from first principles, we use DFT+eDMFT method~\cite{webpage, JPSJ_Haule}. 
Here the downfolding to Hubbard-type models is avoided, and the dynamic correlations are added only to the very localized $3d$ orbitals, which are solution of the Schroedinger equation inside a sphere centered on Mn ion (radius 1.93 bohr). The double-counting between the DFT and DMFT correlations is treated within a nominal scheme, where the double counting is given by U(n$_f^0$-$\frac{1}{2}$)-$\frac{J}{2}$(n$_f^0$-1), with n$_f^0$ being the nominal occupancy of the d orbital  ~\cite{exactDC}. The Coulomb repulsion on these quasi-atomic orbitals used in this work has the Slater form in its fully rotational invariant form with its values estimated by constrained-eDMFT to be $U=10\,$eV, $J=1.18\,$eV.

All crystal structures, except the high temperature orthorhombic structure of BiMnO$_3$, were experimentally determined~\cite{LaMnO3_CS_glp_PhysRevB.57.R3189,MM_CS_BiMnO3_glp_JACS,BiMnO3_HT_LP2_glp_B818645F}, and we checked that they agree very well with the theoretically determined optimized crystal structures (see online materials B to E).  We note that the force in our approach is calculated from the derivative of the free energy, rather than total energy~\cite{Forces_DMFT_glp_PhysRevB.94.195146}, therefore it contains important effects of electronic entropy at finite temperature, which we find to be essential to describe orbitally disordered state in the orthorombic structure, with finite but small orthorhombic distortions.  We also note that the high-temperature orthorombic BiMnO$_3$ crystal structure has not been determined before, and here is predicted for the first time (see online materials A). It would be desirable to confirm it experimentally. 

A schematic view of the room temperature (RT) crystal structure of LaMnO$_3$ and BiMnO$_3$ compounds is shown in Fig.~\ref{fig2}. It consist of one and two inequivalent corner-sharing MnO$_6$ octahedra in LaMnO$_3$ and BiMnO$_3$, respectively. The two inequivalent octahedra, denoted by Mn$_1$O$_6$ and Mn$_2$O$_6$, appear due to the lone-pair of electrons derived from Bi $6s$ orbital~\cite{LP_JT_2_glp_theory_10.1021_cm010090m}, whose direction is shown schematically by the gray arrow in Fig.~\ref{fig2}d. Note that the lone-pair consist of electronic charge, occupying the space between the center of Bi and Mn$_2$ ion~\cite{LP_JT_2_glp_theory_10.1021_cm010090m}, which we mark by the gray arrow.

To describe the Mn 3$d$ orbitals, we use the local coordinate system attached to each Mn ion, in which the hybridization of each orbital with environment is minimal. As expected for octahedral environment, there are three $t_{2g}$ and two $e_g$ orbitals pointing towards the faces and corners of the octahedra, respectively. As shown in Fig.~\ref{fig2}d, the lone-pairs are pointing towards the octahedron face of Mn$_2$ ions, thus for this ion type the $t_{2g}$ orbitals are expected to hybridize with the lone-pair strongly, while the orbitals of Mn$_1$ ions type, should not directly feel the presence of lone-pair in BiMnO$_3$.

{
\centering
\captionof{table}{
Occupancy of the Mn orbitals $(n_{e_{g1}},n_{e_{g2}})$ where $\ket{e_{g1}}\approx \ket{3z^2-1}$ and $\ket{e_{g2}}\approx \ket{x^2-y^2}$.  Note that 
$n_{t_{2g1}} \approx n_{t_{2g2}} \approx n_{t_{2g3}} \approx 1$. The  realized electronic states are
\textit{band-Mott insulator} (BMI), \textit{orbital-selective Mott} (OSM) state, and \textit{site-orbital-selective Mott} (SOSM) state.
}
\label{tab:struct}
 \begin{tabular}{|l|c|c|c|c}
\hline
$\bold{LaMnO_3}$          & T=300K & T=573K  & T=798K \\
\hline
Mn      & ( 0.98, 0.45 ) & ( 0.97, 0.46 ) &  ( 0.76, 0.66 )\\
\hline
state & BMI & BMI & OSM\\
\hline
\hline
$\bold{BiMnO_3}$           & T=300K & T=550K  & T=760K \\
\hline
Mn$_1$   & ( 0.98, 0.44 ) & ( 0.93, 0.47 ) & ( 0.72, 0.69 ) \\
Mn$_2$       & ( 0.98, 0.44 ) & (0.73,  0.69 ) & ( 0.72, 0.69 ) \\
\hline
state & BMI & SOSM & OSM\\
\hline
\end{tabular}
\par
\bigskip
}

\begin{figure*}[bht]
\includegraphics[width=0.95\linewidth]{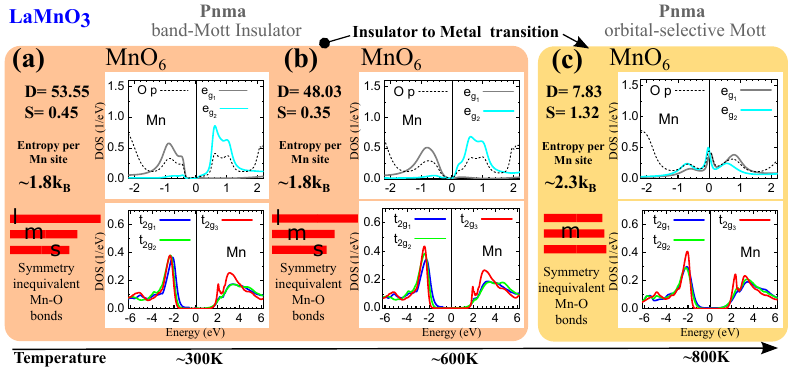}
\caption{(Color online)  \textbf{Temperature dependent electronic states in LaMnO$_3$:}  
Panels (a) to (c) show for each inequivalent Mn octahedra the following properties: (1) values of the distortion indices, $\bold{D}$ and $\bold{S}$; (2) electronic entropy per Mn site; (3) schematic view of the symmetry inequivalent Mn-O bonds represented by red horizontal bars (the horizontal size of the bar is proportional with the bond length, short-s, medium-m, long-l; in panel (c) the three bonds are slightly distorted around the average bond length ); (4) $e_g$ and $t_{2g}$ projected density of states per formula unit versus energy in eV, for each electronic phase at a given temperature.  On top of each panel we give the crystallographic symmetry at a given temperature and a name characterizing the electronic properties of the predicted electronic phases. The panels are connected by arrows suggesting the order parameter of the phase transitions.
}
\label{fig1}
\end{figure*}

While the formal valence of Mn ion is Mn$^{3+}$, our calculation show that due to covalency there is approximately $n_d\approx 4.4$ electrons on each Mn ion (Mn$^{2.6+}$), of which almost exactly three electrons are in the $t_{2g}$ shell, and therefore $n_{e_g}\approx 1.4$. This is in good agreement with valence determined from XAS in Ref.~\onlinecite{Xray2m} ($n_d\approx 4.5$), but it is in stark contrast with the earlier DMFT works in which the number of $e_{g}$ electrons has been fixed to one electron per Mn. 

The degree of the orbital order is usually measured by occupancy ratio $n_{e_{g1}}/(n_{e_{g1}}+n_{e_{g2}})\equiv (c_2)^2$, which was estimated by neutron scattering to be $c_2\approx 0.81$ at room temperature~\cite{LaMnO3_CS_glp_PhysRevB.57.R3189}. Table~1 gives the temperature dependence of the occupancies, as found by eDMFT method. We note that the maximal orbital ordering is for $n_{e_{g1}}\approx 1$, but due to covalency $n_{e_{g2}}$ does not vanish in this limit, hence $c_2$ always remains smaller than 1. In LaMnO$_3$ we get $c_2 = 0.826$, in good agreement with estimates from Ref.~\onlinecite{LaMnO3_CS_glp_PhysRevB.57.R3189}.
In LaMnO$_3$ the orbital polarization changes very little as long as the system has sizable charge gap, while orbital polarization becomes very small in the bad-metal phase above $T=798\,$K. Since the two $e_{g}$ orbitals are not equivalent in the high-temperature structure, there remains a small difference in the occupancy as long as the system is in the orthorombic phase. 
We note that in the earlier DMFT work~\cite{Pavarini} such agreement with experiment could not be achieved, and it was assigned to negligence of the electron-phonon coupling. In the minimal Hubbard model the insulating state must have integer occupancy, hence the constrained $n_{e_{g1}}+n_{e_{g2}}=1$ leads to $c_2\approx 1$ when polarization is large  ($n_{e_{g1}}\approx 1$). Hence we now understand that the earlier disagreement with experiment was due to negligence of Mn covalency.

In BiMnO$_3$ there is an intermediate phase, in which electronic occupancies on the two Mn atoms are very different. Mn$_1$ keeps orbital polarization close to its room temperature value, while Mn$_2$ looses its polarization in this temperature range. As we will see later from structural optimization, such results is hard to guess from crystal structure alone, as the Mn$_2$ octahedra remain quite distorted in this intermediate phase. 

If Fig.~\ref{fig1} we show temperature evolution of the density of states (DOS) for both $t_{2g}$ and $e_g$ orbitals. We also display structural information of the experimental structure. Each octahedra has three different types of Mn-O bond-lengths, which are displayed as red horizontal bars. We label them as short (\textbf{s}$\sim$1.9), medium (\textbf{m}$\sim$2.0) and long (\textbf{l}$\gtrsim$2.1). 
We also provide distortion parameters for octahedra~\cite{Vesta_Momma_db5098}, namely, the \textit{bond distortion index} $\textbf{D}=1000 \sum\limits_{i=1}^6 |l_i-l_{av}|/(6\, l_{av})$
 and the \textit{bond angle variance} $\textbf{S}=\sum\limits_{i=1}^8 (\phi_i-90^o)^2/7$.
 Here $l_i$ and $\phi_i$ are the bond distances and bond-angles, respectively. $l_{av}$ is the average bond length. Compared to Ref. ~\cite{Vesta_Momma_db5098}, we use a formula for $\textbf{D}$ with a multiplying factor of 1000. $\textbf{D}$ and $\textbf{S}$ are zero for an ideal  octahedra.

\begin{figure*}[bht]
\includegraphics[width=0.95\linewidth]{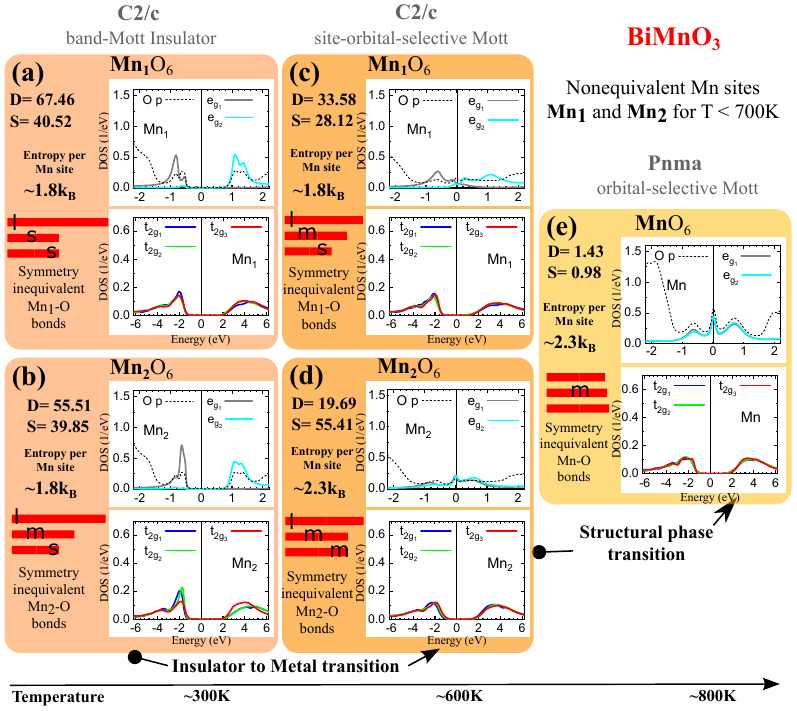}
\caption{(Color online)  \textbf{Temperature dependent electronic states in BiMnO$_3$:}  
Panels (a) to (e) show for each inequivalent Mn octahedra the following properties: (1) values of the distortion indices, $\bold{D}$ and $\bold{S}$; (2) electronic entropy per Mn site; (3) schematic view of the symmetry inequivalent Mn-O bonds represented by red horizontal bars (the horizontal size of the bar is proportional with the bond length, short-s, medium-m, long-l; in panel (e) the three bonds are slightly distorted around the average bond length ); (4) $e_g$ and $t_{2g}$ projected density of states per formula unit versus energy in eV, for each electronic phase at a given temperature. On top of each panel we give the crystallographic symmetry at a given temperature and a name characterizing the electronic properties of the predicted electronic phases. The panels are connected by arrows suggesting the order parameter of the phase transitions.
}
\label{fig3}
\end{figure*}

\begin{figure*}[hbt]
\includegraphics[width=0.95\linewidth]{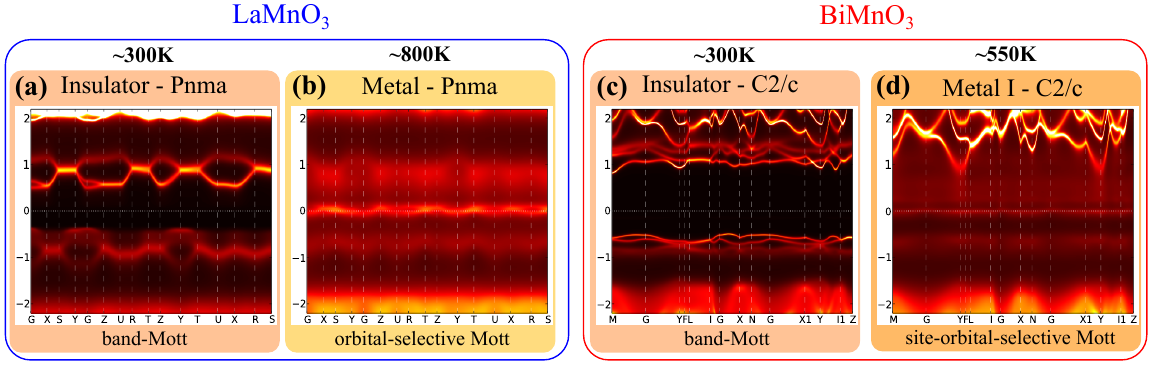}
\caption{(Color online)  \textbf{Spectral functions for various novel phases in (Bi,La)MnO$_3$:} For both compounds we give the spectral functions showing the metal to insulator phase transitions. In panels (a) and (b) we show the spectral functions for the \textit{band-Mott} and \textit{orbital-selective Mott} states for LaMnO$_3$; in panels (c) and (d) we show the spectral functions of the \textit{band-Mott} and \textit{site-orbital-selective Mott} states for BiMnO$_3$.} 
\label{fig4}
\end{figure*}

At all temperatures and in both compounds $t_{2g}$ electrons are in the Mott insulating state, and remain gapped even in the high temperature metallic phase (see Figs.~\ref{fig1} and ~\ref{fig3}). This can be verified by looking at the $t_{2g}$ self-energy, which has a pole inside the gap even when the $e_{g}$ density of states shows metallic behavior (see online materials F to H ). This state is commonly called \textit{orbital-selective Mott state}~\cite{OrbitalSelectiveFirst,orbitalSelective,Hunds_Georges}, and was first introduced in Ref.~\onlinecite{OrbitalSelectiveFirst} to explain the properties of Ca$_{2-x}$Sr$_{x}$RuO$_4$ in which spin-$1/2$ local moments coexist with metallic transport. 
Such a picture of \textit{frozen high-spin} $t_{2g}$ electrons in LaMnO$_3$ was postulated already in the very early works on the double-exchange model~\cite{Zener,Anderson,deGennes}, and was also assumed in the previous DMFT calculations within the minimal two orbital Hubbard model~\cite{Pavarini,Leonov}.  In the limit of weak correlations among $e_g$ electrons, coexisting with the \textit{high-spin core} $t_{2g}$ states, the DMFT problem maps to the ferromagnetic Kondo model, which was shown in Ref.~\onlinecite{Giant_magnetoresistance_FMKondo} to explain qualitatively the giant magnetoresistance in doped LaMnO$_3$.
Nevertheless, to the best our knowledge, direct DMFT simulations with equal dynamic treatment of both $t_{2g}$ and $e_g$ orbitals was lacking, hence the theoretical photoemission spectra and position of the $t_{2g}$ states was unknown. Moreover, as we will see in BiMnO$_3$ the approximation of $t_{2g}$ electrons as \textit{frozen high-spin core} is not really adequate.

Fig.~\ref{fig1}a shows that the room temperature gap is between $e_g$ states, in which the occupied part is strongly admixed with oxygen states.  The orbital with larger occupation ($e_{g1}$) points along the longest Mn-O bond, while the less occupied orbital ($e_{g2}$) points towards the shorter bonds.
The gap between $e_g$'s is not of Mott type, hence we call this state \textit{band-Mott insulator}.
The peak-to peak splitting between these low-energy states is between $1.8-2\,$eV in the orbital ordered state, and the gap collapses around $800\,$K in the high-temperature orthorombic structure, in which the three Mn-O bonds become very similar, although not strictly equal. Consequently, the two $e_g$ orbitals are close to degenerate, but still different (see Fig.~\ref{fig1}c). The phase-transition around $800\,$K is thus a first order metal-insulator transition, strongly coupled with the structural evolution in which the distortion of the octahedra is strongly reduced, but the symmetry of the crystal structure does not change. This calculation shows not only that the details of the crystal structure and the lattice distortions have large impact on the electronic state, but also that the feedback effect of the electronic correlations on the crystal structure is crucial to properly describe the temperature evolution of the system. To understand whether structural disortions or electronic correlations drive the transition, we also calculate the electronic entropy~\cite{FreeEnergy} per Mn-site at three temperatures as shown in Fig.~\ref{fig1}. While normally the metallic phase has small entropy, and the Mott insulating state carries large entropy, here the situation is very different. In the insulating state we find the entropy to be around $1.8\,k_B$ per Mn-site, while in the bad-metallic state it is increased to $2.3\,k_B$, which corresponds to  $35\,$meV entropy gain across the transition, which is a decisive contribution to the free energy for the metal-insulator transition.

The important consequence of the Mott insulating $t_{2g}$ states coexisting with the metallic $e_{g}$ states at high temperature in the existance of so-called \textit{orbital-selective Mott state} (OSM), in which $e_g$ electrons have enormous scattering rate (of the order of eV), and consequently small mobility so that there are no real bands present in the phtoemission spectra, but rather just a very incoherent spectral weight (see Fig.~\ref{fig4}). Photoemission on LaMnO$_3$ revelead that the system is of charge transfer-type insulator, with strong admixure of oxygen $p$ and Mn-$e_g$ weight around 1eV below the Fermi level~\cite{Xray2m}. In Ref.~\onlinecite{PhotoemissionCheong} a strong peak at approximately $-3\,$eV and a shoulder between $-1.5\,$eV and the fermi level was identified, which we now interpret as the $t_{2g}$ and $e_g$ peak in the density of states, respectively. Even more revealing are optical experiments, which show strong temperature dependent peak around $2\,$eV  present only in the orbitally ordered state~\cite{Optics_LaMnO3}, and replaced by broad distribution of optical weight above $800\,$K. By analizing low temperature data in Ref.~\onlinecite{Optics_LaMnO3_recent} it was found that the $2\,$eV peak in optics comes from $e_g-e_g$ transitions, and that $t_{2g}-e_g$ transitions appear at $3.8\,$eV. This is all consistent with our spectra in Fig.~\ref{fig1}a, as the $e_g-e_g$ distance is close to $2\,$eV and the $t_{2g}-e_g$ distance is close to $4\,$eV at room temperature, while in disordered state above $800\,$K  the gap dissapears and the scattering rate broadens low energy excitations. We note that in earlier DMFT works, the $e_g-e_g$ splitting in the ordered state was somewhat too large ($\approx 3.5\,$eV~\cite{Pavarini} and $\approx \,4$eV~\cite{Leonov}), pointing again to the shortcomings of the low energy Hubbard-type modeling.
Finally we note that theoretical DOS in Fig.~\ref{fig1} are plotted using experimental crystal structures, but the results do not change appreciably when the theoretical crystal structures are used. 


While large amount of experimental data is available for LaMnO$_3$, very little is known about BiMnO$_3$. Our calculations (Fig.~\ref{fig3}) show strong similarity with LaMnO$_3$, but also some important differences. 

In particular, there is stronger anisotropy within the DOS of the $t_{2g}$ and $e_g$ orbitals for the Mn$_2$-ion compared to the Mn$_1$-ion (see Fig.~\ref{fig3}a and b). For example, there is a very sharp peak in the DOS of two $t_{2g}$ orbitals at $-2.0\,$eV and $e_g$ orbital at $-0.6\,$eV, corresponding to  Mn$_2$-ion (Fig.~\ref{fig3}b). This is a consequence of the anisotropic hybridization between the lone-pair, O-p and Mn$_2$-d orbitals.  We recall that the lone-pair of Bi-$6s$ state points towards the face of the distorted Mn$_2$O$_6$ octahedra as shown in Fig.~\ref{fig2}d, thus this anisotropic hybridization is present in the system (see online materials I). The anisotropic hybridization of the $t_{2g}$ orbitals with the lone-pair in the case of Mn$_2$  atoms is also confirmed by the fact that the Mn$_1$ atoms have similar distortions, but the corresponding DOS of the $t_{2g}$ orbitals is isotropic (Fig.~\ref{fig3}a). The highly distorted monoclinic crystal structure of BiMnO$_3$ compound, is a consequence of the competition between the Jahn-Teller octahedra distortions that are squeezing the lone-pair of Bi-$6s$ state. This is different from the case of LaMnO$_3$ compound, where the lone-pair is absent, thus the Jahn-Teller octahedra distortions appears in the orthorhombic crystal structure. 

%
\begin{figure*}[hbt]
\includegraphics[width=0.95\linewidth]{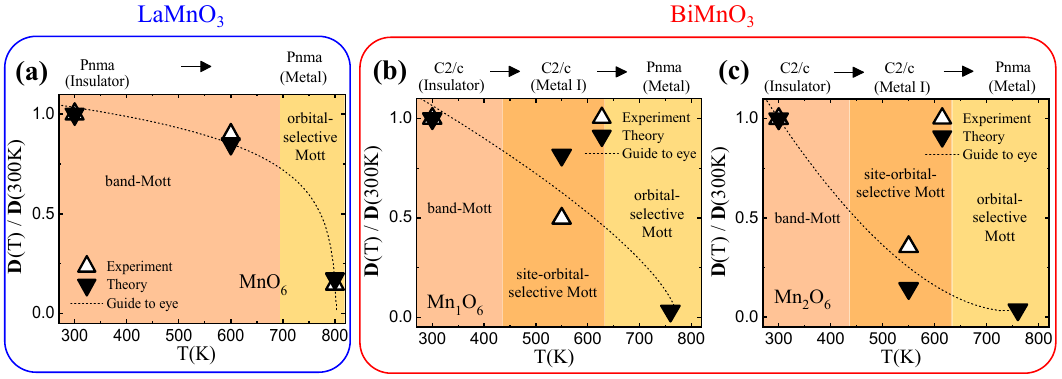}
\caption{(Color online)  \textbf{Structural relaxations within LDA+eDMFT for all the temperature dependent novel electronic phases in (Bi,La)MnO$_3$:} In each panel we plot the Bond Distortion ($\bold{D}$) index versus temperature, normalized to its values at room temperature. In panel (a) we plot $\bold{D}$ for the Mn site in LaMnO$_3$. In panels (b) and (c) we plot $\bold{D}$ for the two inequivalent Mn sites in BiMnO$_3$. Above each panel we give the temperature dependent crystal symmetry and the properties of that electronic state (metal or insulator). The arrows point to the type of phase transition (structural or metal-insulator phase transition).
} 
\label{fig5}
\end{figure*}

At room temperature, the local Jahn-Teller distortions for both inequivalent Mn ions are large enough to decrease the e$_{g1}$ hybridization with the O-p states allowing localization of $e_g$ electrons, even though the e$_g$ orbitals also hybridize with the lone-pair. These local distortions are depicted by red segments and given in terms of distortion indices in Fig.~\ref{fig3}a and b.

Note that larger values of the bond distortion index $\bold{D}$ in octahedra corresponds to more anisotropic hybridization between the two e$_g$ orbitals and O-p states, causing orbital polarization and consequently localization of electrons.  With increasing temperature our theoretical method reveals that the local Jahn-Teller distortions decrease (meaning smaller $\bold{D}$ values), thus the  e$_{g1}$ hybridization with the O-p states is increasing while the e$_{g2}$ hybridization is decreasing (see online materials J). At the same time the lone-pair is exerting pressure on the neighboring ions to gain more space, thus it is contributing further to decreasing local distortions. Since the lone-pair points towards the face of the Mn$_2$O$_6$ distorted octahedra, the local distortions of the Mn$_2$ site get reduced faster with increasing temperature then for the Mn$_1$ site. These local distortions are depicted by red segments and given in terms of distortion indices in Fig.~\ref{fig3}c and d. Consequently, when the local distortions are small enough, e$_{g1}$ and e$_{g2}$ hybridization with the O-p states become similar, thus the $e_g$ gap on Mn$_2$ ion collapses, and the system becomes bad metal.  While decreasing the local distortions, the anisotropic hybridization of the $t_{2g}$ with the lone-pair and O-p orbitals disappears, i.e., the resonance in $t_{2g}$ DOS dissapears going from Fig.~\ref{fig3}b to d. However, Mn$_1$ ion remains close to its low-temperature state, for one thing, its orbital occupancy in Table~1 remains unchanged, and the $e_g$ states show DOS with a pseudogap (Fig.~\ref{fig3}c). Due to the qualitative difference in physical state on the two Mn ions, we call this state \textit{site-orbital-selective Mott} state, and we note that the orbital order remains strong only on Mn$_1$ ion. Such partial orbital order on half of Mn ions is consistent with RXS experiments~\cite{LaMnO3_RXS_PhysRevB.73.224112}, in which the room temperature signal was roughly factor of two stronger than in the intermediate state. Surprisingly, the entropy on Mn atoms is very weakly  temperature dependent between the room temperature and the first transition (around 600$\,$K), at which the entropy on Mn$_2$ increases for $0.5\,k_B$, while the entropy on Mn$_1$ remains almost unchanged. Only at the second transition around 800$\,$K the entropy of Mn$_1$ increases to its saturated value within the bad metallic state.

We also note that the electronic spectral weight of the $t_{2g}$ orbitals pointing towards the lone-pair is strongly temperature dependent, therefore treating $t_{2g}$ electrons as a fixed core spin would not result in correct prediction of the crystal structures.

Fig.~\ref{fig4} displays the single-particle spectral functions $A(\vk,\omega)$ in the insulating room temperature state, and in the first metallic state. The room temperature direct gap in LaMnO$_3$ and BiMnO$_3$ are $\approx 0.8\,$eV and $1.2\,$eV, which are in decent agreement with optical experiments ( $0.9\,$eV in LaMnO$_3$~\cite{Optics_LaMnO3} and $1.2\,$eV in BiMnO$_3$~\cite{gap1_BiMnO3_10.1063_1.3457786,gap2_BiMnO3_10.1063_1.3427499}). We note that the prediction of the gap size in BiMnO$_3$ seems to be very challenging for other ab-initio methods, as the gap size is either too large (hybrid functional $1.7\,$eV~\cite{gapTHeory4_BiMnO3_PhysRevB.91.184113}) or too small (modified Becke-Johnson exchange functional  $0.75\,$eV~\cite{gapTHeory5_BiMnO3_ZHU201665}; $0.5\,$eV or less by various LDA+U calculations
\cite{gapTHeory3_BiMnO3_0953-8984-22-29-295404,gap3_BiMnO3_PhysRevB.81.144103,gapTHeory2_BiMnO3_PhysRevB.78.104106,gapTHeory1_BiMnO3_0953-8984-16-48-026}).
Finally, the \textit{orbital-selective Mott} phase, in which the orbital order is melted away, is displayed in Fig.~\ref{fig4}b. The spectral functions has no sharp bands, only the incoherent spectral weight, hence the charge mobility is very low. This state is also realized in the ferromagnetic Kondo model of Ref.~\onlinecite{Giant_magnetoresistance_FMKondo}, therefore it might show giant magnetoresistance.
 Fig.~\ref{fig4}d shows the new phase \textit{site-orbital-selective Mott} state, realized in the intermediate temperature range where Mn$_1$ shows orbital ordering and Mn$_2$  does not. This state is a superposition of very incoherent \textit{orbital-selective Mott}  phase and the insulating phase, which shows somewhat more sharp bands above $1\,$eV.

Fig.~\ref{fig5} shows the comparison between the theoretical and experimental temperature dependence of the distortion index $\textbf{D}$. In LaMnO$_3$ it changes only for $20\%$ between the room temperature and $550\,$K, while it collapses abruptly around $800\,$K. We note an excellent agreement between the theoretical and experimental distortion index. In BiMnO$_3$ the two Mn sites show very different behaviour: the changes on Mn$_1$ are comparable to LaMnO$_3$ ($20\,\%$), while on Mn$_2$ ion the distortion is almost complete at $550\,
$K ($80\,\%$). In the high temperature phase the distortion is small but finite. The reason for such markedly different behaviour of the two sites is that Mn$_2$ ion hybridizes with the Bi 6s lone-pair, while Mn$_1$ does not.
Finally, we note that the measured and theoretical distortion index have similar trend, except that the disproportionation between Mn$_1$ and Mn$_2$ seems a bit weaker in experiment than in theory.

In conclusion, using  eDMFT methods we found novel electronic state in Mn perovskite system, such as \textit{site-orbital-selective Mott} state, which hosts Mott insulating $t_{2g}$ states, and a combination of semi-metallic and metallic $e_g$ states on two inequivalent Mn atoms. 
We showed that the isostructural transition in BiMnO$_3$ at $474\,$K should be characterized as the insulator-metal transition at which half of Mn atoms give up the orbital order, while the fully disordered state is reached only at the structural transition of $768\,$K. We showed that
the electronic entropy plays the decisive role in the metal-insulator transition in both compounds.
By self-consistent determination of both the correlated electronic state and the corresponding relaxed crystal structure we showed that the temperature evolution of phases in LaMnO$_3$ and BiMnO$_3$ can be properly explained from first principles, thus demonstrating the predictive power of the eDMFT method for structural and electronic properties. In addition, we predict the temperature dependence of the spectral functions for the novel phases found in Mn perovskite systems, predictions that can be tested experimentally by ARPES.

\section*{Acknowledgements}
G.L.P. was supported by the U.S. Department of Energy, Office of Science, Basic Energy Sciences, as a part of the Computational Materials Science Program, funded by the U.S. Department of Energy, Office of Science, Basic Energy Sciences, Materials Sciences and Engineering Division while at Rugers and by a grant of the Romanian Ministry of Education and Research, CNCS-UEFISCDI, project number PN-III-P1-1.1-TE-2019-1767, within PNCDI III while at USV . K.H. was supported by the NSF DMR-1709229.

\section*{Author contributions}
G.L.P. carried out the calculations. K.H. developed the code. G.L.P and K.H. analyzed the results and wrote the paper. 

\smallskip
\smallskip
\smallskip
\section*{Competing financial interests} 
The authors declare no competing interests.

\bibliographystyle{naturemag}
\bibliography{Ref_Paper_PRB_AMnO3_GLP_KH}

\end{document}